\documentclass[11pt]{article}
\usepackage{amsmath}
\usepackage{amsfonts}
\newtheorem{theorem}{Proposition}[section]
\newcounter{mnotecount}[section]

 \renewcommand{\themnotecount}{\thesection.\arabic{mnotecount}}

 \newcommand{\mnote}[1]
 {\protect{\stepcounter{mnotecount}}$^{\mbox{\footnotesize
 $
 \bullet$\themnotecount}}$ \marginpar{
 \raggedright\tiny\em
 $\!\!\!\!\!\!\,\bullet$\themnotecount: #1} }

\def\be{\begin{equation}}
\def\ee{\end{equation}}
\def\bea{\begin{eqnarray}}
\def\eea{\end{eqnarray}}
\def\bea*{\begin{eqnarray*}}
\def\eea*{\end{eqnarray*}}

\begin{document}
\title{The problem of characterising stationary data for the vacuum Einstein equations}
\author{Paul Tod,\\Mathematical Institute\\and\\St John's College,\\
Oxford}

\maketitle
\begin{abstract}
We take a step towards characterising stationary data for the vacuum Einstein equations, by finding a necessary condition on initial data for which the evolution is a solution of the vacuum equations 
admitting a Killing vector, which is time-like at least in some region of the Cauchy development. 
\end{abstract}
\section{Introduction: the Problem}
In this article we consider the problem of characterising stationary data for the vacuum Einstein equations, that is to say of identifying data for 
which the evolution is a solution of the vacuum equations 
admitting a Killing vector which is time-like at least in some region of the Cauchy development. By data here we mean a 3-manifold $\Sigma$ with a Riemannian metric $g_{ij}$ and 
a symmetric tensor $K_{ij}$, understood as the second fundamental form of $\Sigma$, together satisfying the vacuum constraint equations (\ref{c1})--(\ref{c2}) below. We obtain a necessary condition 
expressed in terms of the data, and a simple interpretation of it in terms of the associated space-time.

A particular case of this problem, when $K_{ij}=0$, was considered in \cite{bt}. Necessary conditions were found for staticity, and the problem was completely solved in the case when the Ricci tensor 
of $g_{ij}$ has distinct eigenvalues\footnote{If this Ricci tensor has a repeated eigenvalue then the space-time is type D, and all static vacuum type D solutions are known explicitly, \cite{ExSo}, so the 
problem is also solved in this case.}. 
We now consider the general case when $K_{ij}\neq 0$. Note that the method detects the presence of space-time Killing vectors and it is then a secondary question whether the Killing vector is time-like in a 
suitable region. Thus we shall be addressing the larger question of data leading to space-time symmetries.

We consider data subject to the vacuum constraint equations:
\be\label{c1}R+K^2-K_{ij}K^{ij}=0,\ee
\be\label{c2}D_iK^i_{\;\;j}-D_jK=0,\ee
where $D_i$ is the Levi-Civita covariant derivative, $R$ the Ricci scalar of $\Sigma$ and $K=g^{ij}K_{ij}$.

Beig and Chru\'sciel \cite{bc} introduced the notion of Killing Initial Data or KIDs as follows: if the space-time $M$ which evolves from this data admits a 
Killing vector $X$ then $X$ can be decomposed with respect to the normal to $\Sigma$ as a pair $(N,Y^i)$ consisting of a function $N$ on $\Sigma$ and a vector field 
$Y^i$ tangent to $\Sigma$ and, by virtue of the Killing equations in $M$, this pair satisfies the pair of equations:
\begin{eqnarray}
D_{(i}Y_{j)}&=&-NK_{ij}\label{k1}\\
D_iD_jN+\mathcal{L}_YK_{ij}&=&N(R_{ij}+KK_{ij}-2K_{im}K_j^{\;m}).\label{k2}
\end{eqnarray}
Further, given the data and the KID $(N,Y^i)$, the same authors introduce the \emph{Killing development}, \cite{bc}, which is the 4-metric:
\be\label{k3}
g=g_{ij}(dx^i+Y^idu)(dx^j+Y^jdu)-N^2du^2,
\ee
and is a solution of the vacuum field equations admitting a Killing vector $X=\partial/\partial u$ and inducing the original data at $\Sigma=\{u=0\}$. The Killing vector 
is evidently time-like, space-like or null according to the sign of $N^2-g_{ij}Y^iY^j$.

Beig et al \cite{bcs} show that KIDs are non-generic, in the sense that generic sets of data do not admit them, they prove the existence of data with none, and they give some
necessary conditions on data for non-existence of a KID.

Our interest here is in the conditions for non-existence, and we shall obtain sufficient conditions which are simpler than those in \cite{bcs}, and with a 
direct geometric interpretation. The results are contained in the following proposition:
\begin{theorem}

 \begin{enumerate}
 
  \item There can be no KIDs if the determinant of a certain $10\times 10$-matrix $A$ is nonzero. The entries of $A$ are polynomials in the tensors $R_{ij},D_iR_{jk},K_{ij},D_iK_{jk}$ 
  and $D_iD_jK_{km}$ where $R_{ij}$ is the Ricci tensor of $g_{ij}$.
  \item Using $3+1$-formalism, the entries of $A$ can be expressed directly in terms of the Weyl tensor of the space-time evolving from the data, and the determinant 
  condition in part 1 can then be written
  \be\label{j1}
dI\wedge dJ\wedge d\overline{I}\wedge d\overline{J}\neq 0.
\ee
  where $I$ and $J$ are the two complex scalar invariants of the Weyl spinor defined in (\ref{i11}) below.  
 \end{enumerate} 
\end{theorem}

If the determinant is zero, then there will be candidate KIDs, and this will always be the case for example with data for algebraically special space-times when 
necessarily $I^3-6J^2$ vanishes. To see whether the candidate KIDs are actually KIDs then requires derivatives of the constraints given by $A$, and we won't pursue this question here. 

\medskip

\medskip

The method of proof is to prolong the system (\ref{k1})--(\ref{k2}) and obtain a 
first-order linear system for a larger set of variables (see \cite{bceg} and \cite{t1} for earlier applications of the method). This linear system can be regarded 
as covariant constancy for a section of a particular vector bundle with connection, and existence of solutions can then be analysed in terms of the curvature of the 
connection. In Section 2 we find a sufficient condition for non-existence of a KID, equivalently a condition that the data not be data for a stationary space-time\footnote{Or 
indeed a space-time admitting any Killing vector.}, in terms of the 
non-vanishing of a determinant expressed in terms of the data and its derivatives. In Section 3, we express this condition in space-time terms, as in (\ref{j1}), when it is 
more perspicuous.


\medskip

We shall adopt the convention that indices from the start of the alphabet, $a,b,c,\ldots$ are 4-dimensional indices, that is indices on tensors in the space-time $M$, and indices 
from the middle of the alphabet, $i,j,k,\ldots$ are 3-dimensional indices, on tensors tangent to $\Sigma$. With $n_a$ as the unit time-like 
normal to $\Sigma$ in $M$ we therefore have 
\[N=X^an_a,\;\;Y^i=\Pi^i_aX^a,\]
where $\Pi^i_a$ is the projection into $\Sigma$. To simplify equations, but with a slight abuse of notation, we shall usually 
omit the projection tensors $\Pi_a^i$ and $\Pi_i^a$.

\section{The bundle and connection}
We begin by prolonging (\ref{k1})--(\ref{k2}). Introduce new variables $V_i:=D_iN$ and $M_{ij}:=D_{[i}Y_{j]}$ so that
\begin{eqnarray}
D_iN&=&V_i\label{s1}\\
D_iY_j&=&-NK_{ij}+M_{ij}\label{s2}\\
D_iV_j&=&N(R_{ij}+KK_{ij})-Y^kD_kK_{ij}-K_{im}M_j^{\;m}-K_{jm}M_i^{\;m}.\label{s3}
\end{eqnarray}
We need an equation for $D_iM_{jk}$. Commute derivatives on $Y_j$ and rearrange to find
\[D_kM_{ij}=-2V_{[i}K_{j]k}-2ND_{[i}K_{j]k}-R_{ijk\ell}Y^\ell,\]
with the convention
\[(D_iD_j-D_jD_i)Z_k=R_{ijk}^{\;\;\;\;\ell}Z_\ell.\]
Then by relabelling and reordering
\be\label{s4}
D_iM_{jk}=-2ND_{[j}K_{k]i}-R_{jki\ell}Y^\ell-2V_{[j}K_{k]i},
\ee
and we have the desired linear system (\ref{s1})--(\ref{s4}). 

To see this system as a connection on a vector bundle, introduce the ten component column-vector  
$\Psi_\alpha=(N,Y_j,V_j,M_{jk})^T$, which is a section of the vector bundle $\Lambda^0(\Sigma)\oplus\Lambda^1(\Sigma)\oplus\Lambda^1(\Sigma)\oplus\Lambda^2(\Sigma)$, 
when the system appears as
\be\label{s5}\mathcal{D}_i\Psi_\alpha:=D_i\Psi_\alpha-\Gamma_{i\alpha}^{\;\;\;\;\beta}\Psi_\beta=0,\ee
with $\Psi_\beta=(N,Y_p,V_p,M_{pq})^T$ and
\be\label{g1}
\Gamma_{i\alpha}^{\beta}=\left(\begin{array}{cccc}
           0 & 0 & \delta_i^p &0 \\
          -K_{ij} & 0 & 0 &  \delta_i^{\;p}\delta_j^{\;q}\\
           R_{ij}+KK_{ij} & -D^pK_{ij} & 0 & -K_i^{\;p}\delta_j^{\;q}-K_i^{\;q}\delta_j^{\;p}\\
           -2D_{[j}K_{k]i} & -R_{jki}^{\;\;\;\;p} &-2\delta_{[j}^{\;p}K_{k]i} & 0\\
\end{array}\right).
\ee
Note that $\Psi$ contains all the components of $X^a$, namely $N$ and $Y^i$, and all the components of $\nabla_aX_b$ since, as one may 
calculate,
\be\label{d1}
\nabla_aX_b=M_{ab}+\tilde{V}_an_b-n_a\tilde{V}_b,\ee
where
\[\tilde{V}_a=V_a+K_{ab}Y^b,\]
so that $V_i$ and $M_{ij}$ with $Y^i$ and the data determine all components of the derivative of $X$.

We won't need it but for completeness note
\[X^a=Nn^a+Y^a,\]
with
\[n^aN_{,a}=-Y^aA_a,\]
and
\[P_a^{\;d}n^c\nabla_cY_d=\tilde{V}_a-NA_a\]
where $P_a^{\;b}$ is projection orthogonal to $n^a$ and $A^a$ is the acceleration of the normal congruence i.e.
\[\nabla_an_b=-n_aA_b+K_{ab}.\]
The acceleration is typically fixed by a gauge condition but, for our purposes, we don't need to do that.

\medskip

The usual argument can now be followed: a solution of the prolonged system corresponds to a constant section of the vector bundle in the connection $\mathcal{D}$ 
and so must annihilate the 
curvature tensor of $\mathcal{D}$; in particular the rank of this curvature must be less than maximal, and this will be the source of the desired conditions 
on the data.

\medskip

Rather than writing out the curvature tensor, we note that commutators on $N$ and $V_j$ have been seen to give identities. New conditions arise only 
from the commutator of Levi-Civita 
derivatives on $V_j$ and $M_{jk}$ and for these we find:
\begin{itemize}
\item
By considering $\epsilon^{ij}_{\;\;\;m}D_iD_jV_n$ and using (\ref{s1})--(\ref{s4}) we deduce the vanishing of
\[C^{(1)}_{mn}=N\epsilon^{ij}_{\;\;\;m}(D_iR_{jn}+K_{,i}K_{jn}+KD_iK_{jn}+2K_i^{\;p}D_pK_{jn}+K_j^{\;p}D_nK_{pi}+K_n^{\;p}D_jK_{pi})\]
\[-\epsilon^{ij}_{\;\;\;m}Y^pD_pD_iK_{jn}\]
\[+\epsilon^{ij}_{\;\;\;m}(V_i(R_{jn}+KK_{jn})-V_pK_j^{\;p}K_{ni}+K_n^{\;p}K_{pi}V_j)+\epsilon^{ij}_{\;\;\;n}G_{jm}V_i\]
\be\label{con1}-\epsilon^{ij}_{\;\;\;m}(M_j^{\;p}D_iK_{np}+M_n^{\;p}D_iK_{jp}+M_i^{\;p}D_pK_{jn}).\ee
Here we've used the identity, valid in three dimensions,
\[R_{ijk\ell}=-\epsilon_{ij}^{\;\;\;p}\epsilon_{k\ell}^{\;\;\;q}G_{pq},\]
and $G_{ij}$ is the Einstein tensor.

Note $C^{(1)}_{mn}g^{mn}=0=\epsilon^{mnr}C^{(1)}_{mn}$ identically, using both  constraints, so that $C^{(1)}_{mn}$ is symmetric and trace-free and its vanishing can be 
expected to impose five linear conditions on $\Psi_\alpha$.
\item By considering $\epsilon^{ij}_{\;\;\;m}\epsilon^{pq}_{\;\;\;n}D_iD_jM_{pq}$ and using (\ref{s1})--(\ref{s4}) we deduce the vanishing of
\[C^{(2)}_{mn}=N\epsilon^{ij}_{\;\;\;m}\epsilon^{pq}_{\;\;\;n}(-2D_iD_pK_{qj}-2K_{qj}(R_{ip}+KK_{ip}))-2N(G_{in}K^i_{\;m}-KG_{mn})\]
\[+2\epsilon^{ij}_{\;\;\;m}\epsilon^{pq}_{\;\;\;n}K_{qj}Y^kD_kK_{ip}-2Y^iD_iG_{mn}\]
\[-\epsilon^{ij}_{\;\;\;m}\epsilon^{pq}_{\;\;\;n}(2V_pD_iK_{qj}+2V_iD_pK_{qj})\]
\be\label{con2}+2\epsilon^{ij}_{\;\;\;m}\epsilon^{pq}_{\;\;\;n}K_{qj}(K_{is}M_p^{\;s}+K_{ps}M_i^{\;s})+2G_{mi}M^i_{\;n}+2G_{ni}M^i_{\;m},\ee
which is again symmetric and trace-free, so again is expected to give five equations.
\end{itemize}
The two constraints together amount to ten linear equations on the ten-dimensional space of $\Psi_\alpha$ so that there is an implied $10\times 10$-matrix $A$ which we find 
more explicitly below. For non-trivial solutions to exist, this matrix must have zero determinant and the vanishing of the determinant corresponds to the vanishing of a polynomial 
in the tensors $R_{ij},D_iR_{jk},K_{ij},D_iK_{jk}$ and $D_iD_jK_{km}$. This proves part 1 of Proposition 1.1. Similar conditions given 
in \cite{bcs} involve higher orders in derivatives, specifically up to second order on $R_{ij}$ and third order on $K_{ij}$.

\medskip

From the way they are derived, these identities should be expected to be equivalent to the identity obtained as the vanishing of the Lie derivative of the Weyl tensor $C_{abcd}$ along $X$:
\[0={\mathcal{L}}_XC_{abcd}=X^e\nabla_eC_{abcd}+C_{abce}\nabla_dX^e+C_{abed}\nabla_cX^e+C_{aecd}\nabla_bX^e+C_{ebcd}\nabla_aX^e,\]
and this in turn would give ten equations, linear in the variables $\Psi_\alpha$.

\medskip

\medskip

To see that these two sets of identities are in fact equivalent we interpret the two $C^{(i)}_{mn}$ directly in terms of the Weyl tensor and its derivatives by 
following the $(3+1)$-formalism given, for example, by Ellis and van Elst \cite{ev}. With $n^a$ as the future-pointing, unit time-like normal to $\Sigma$, define 
electric and magnetic parts of the Weyl tensor by
\[E_{ab}=C_{acbd}n^cn^d,\;\;B_{ab}=\frac{1}{2}\epsilon_{ac}^{\;\;\;ef}C_{efbd}n^cn^d.\]
Both are trace-free and symmetric tensors tangent to $\Sigma$ and so may be written as $E_{ij}$ and $B_{ij}$ 
(but note that we have chosen the opposite sign on $B_{ij}$ from \cite{ev}) and both are 
expressible in terms of the Cauchy data. These expressions are
\be\label{e1}E_{ij}=R_{ij}+KK_{ij}-K_{ik}K_j^{\;k}\ee
\be\label{b1}B_{ij}=-\epsilon_i^{\;km}D_kK_{mj}\ee
respectively.

We may express the Weyl tensor algebraically in terms of its electric and magnetic parts as
\[C_{abcd}=-4n_{[a}E_{b][c}n_{d]}-\epsilon_{ab}^{\;\;\;\;pq}\epsilon_{cd}^{\;\;\;\;rs}n_pE_{qr}n_s+2\epsilon_{ab}^{\;\;\;\;pq}n_pB_{q[c}n_{d]}+2\epsilon_{cd}^{\;\;\;\;pq}n_{[a}B_{b]p}n_q\]
so that the Weyl tensor, and indeed any tensor with Weyl tensor symmetries, is determined by its electric and magnetic parts. Note also that
\[C_{abcd}n^d=2n_{[a}E_{b]c}-\epsilon_{ab}^{\;\;\;\;pq}n_pB_{qc}.\]
Now we use (\ref{e1}) and (\ref{b1}) in (\ref{con1}) and (\ref{con2}). The constraints become
\[C^{(1)}_{mn}=N(\epsilon_{ij(m}D^iE^j_{\;n)}+\epsilon_m^{\;\;ij}\epsilon_n^{\;\;pq}K_{ip}B_{jq}+K_{i(m}B^i_{\;n)}-KB_{mn})\]
\[+Y^kD_kB_{mn}+2\epsilon_{ij(m}V^iE^j_{\;n)}+2M_{(m}^{\;k}B_{n)k}=0,\]
and, removing a factor 2 for convenience,
\[\frac{1}{2}C^{(2)}_{mn}=N(\epsilon_{ij(m}D^iB^j_{\;n)}-\epsilon_m^{\;\;ij}\epsilon_n^{\;\;pq}K_{ip}E_{jq}-K_{i(m}E^i_{\;n)}+KE_{mn})\]
\[-Y^kD_kE_{mn}+2\epsilon_{ij(m}V^iB^j_{\;n)}+4E_{i(m}M^i_{\;n)}=0.\]
As a check, note the symmetry $E\rightarrow B\rightarrow -E$ relating $C^{(1)}$ and $C^{(2)}$. 

We claim these equations express the vanishing of the electric and magnetic parts of the tensor $Q_{abcd}$ defined by
\[Q_{abcd}:={\mathcal{L}}_XC_{abcd}\]
\[=X^e\nabla_eC_{abcd}+C_{abce}\nabla_dX^e+C_{abed}\nabla_cX^e+C_{aecd}\nabla_bX^e+C_{ebcd}\nabla_aX^e.\]
This is clearly a tensor with Weyl tensor symmetries and so is characterised in turn by its electric and magnetic parts. It is a straightforward matter to put $Q_{abcd}$ into 
the $(3+1)$-formalism and justify the claim. The 
two constraints are therefore equivalent to the vanishing of ${\mathcal{L}}_XC_{abcd}$ expressed in terms of the data. This proves the first half of part 2 of the Proposition. 
The rest is simpler in the two-component spinor formalism, which we turn to next.

\section{Spinor form of the constraints}
We have moved emphasis slightly from consideration of initial data to the question of when does a given 4-dim vacuum solution admit a Killing vector.\footnote{This problem was 
addressed in \cite{nom} and the solution is in principle given there, in Theorem 6.1, but not in a very accessible form.} This question is simpler to answer in 2-component 
spinors when the necessary condition is 
the existence of a nontrivial solution to (\ref{c4}) below, regarded as an equation in $(X^a,\phi_{AB})$, but there is an obvious necessary condition that is easy to obtain. The Weyl spinor has two 
complex invariants $I,J$ given in (\ref{i11}) below, and it's clear that these must be constant along any Killing vector. Thus there can be no Killing vectors if 
these invariants define four functionally-independent real scalars, that is if
\be\label{j11}
dI\wedge dJ\wedge d\overline{I}\wedge d\overline{J}\neq 0.
\ee
This will turn out to be equivalent to the vanishing of $C^{(1)}$ and $C^{(2)}$ as in the previous section, completing the proof of Proposition 1.1.

\medskip

In the 2-component spinor formalism, the definition of the Lie derivative applied to the Weyl spinor $\psi_{ABCD}$ leads to
\be\label{c4}
{\mathcal{L}}_X\psi_{ABCD}:=X^{EE'}\nabla_{EE'}\psi_{ABCD}+4\phi_{(A}^{\;\;\;E}\psi_{BCD)E}=0\ee
where $\phi_{AB}$ is a symmetric spinor obtained from the derivative of the Killing vector via
\[\nabla_{AA'}X_{BB'}=\phi_{AB}\epsilon_{A'B'}+\overline{\phi}_{A'B'}\epsilon_{AB}.\]
The system (\ref{c4}) consists of five complex equations for the four real components of $X$ and the three complex components of $\phi_{AB}$. 
Generically the Weyl spinor is invertible in the sense that there exists a symmetric spinor $\chi_{ABCD}$ with
\[\chi_{AB}^{\;\;\;\;\;\;CD}\psi_{CD}^{\;\;\;\;\;\;EF}=\delta_{(A}^{\;\;\;(E}\delta_{B)}^{\;\;\;F)}.\]
In fact this fails iff $J=0$ as is seen as follows: recall the definition of the complex scalar invariants
\be\label{i11}I=\psi_{ABCD}\psi^{ABCD},\;J=\psi_{ABCD}\psi^{CDEF}\psi_{EF}^{\;\;\;\;\;AB},\ee
then the Cayley-Hamilton Theorem for the Weyl spinor thought of as a $3\times 3$ matrix gives
\[\psi_{AB}^{\;\;\;\;\;EF}\psi_{EF}^{\;\;\;\;\;GH}\psi_{GH}^{\;\;\;\;\;PQ}-\frac{1}{2}I \psi_{AB}^{\;\;\;\;\;PQ}-\frac{1}{3}J\delta_{(A}^{\;\;(P}\delta_{B)}^{\;\;Q)}=0\]
so that
\[\chi_{AB}^{\;\;\;\;\;PQ}=\frac{3}{J}(\psi_{AB}^{\;\;\;\;\;EF}\psi_{EF}^{\;\;\;\;\;PQ}-\frac{1}{2}I\delta_{(A}^{\;\;(P}\delta_{B)}^{\;\;Q)}),\]
which is evidently well-defined where $J\neq 0$.

We introduce the spinor $\chi_{E'ABCDE}$ by
\[\chi_{E'ABCDE}=\nabla_{E'E}\psi_{ABCD},\]
which is then symmetric in the unprimed indices by virtue of the vacuum Bianchi identity. In the NP formalism, by introducing a normalised spinor dyad, 
(\ref{c4}) can be written out 
in components as
\be\label{m1}
\left(\begin{array}{ccccccc}
           \chi_{0'0} & \chi_{1'0} & \chi_{0'1} & \chi_{1'1} & 4\psi_1 & -4\psi_0 & 0 \\
            \chi_{0'1} & \chi_{1'1} & \chi_{0'2} & \chi_{1'2} & 3\psi_2 & -2\psi_1 & -\psi_0 \\
             \chi_{0'2} & \chi_{1'2} & \chi_{0'3} & \chi_{1'3} & 2\psi_3 & 0 & -2\psi_1 \\
              \chi_{0'3} & \chi_{1'3} & \chi_{0'4} & \chi_{1'4} & \psi_4 & 2\psi_3 & -3\psi_2 \\
               \chi_{0'4} & \chi_{1'4} & \chi_{0'5} & \chi_{1'4} & 0 & 4\psi_4 & -4\psi_3 \\

\end{array}\right)
\left(\begin{array}{c}
X^{00'}\\
X^{01'}\\
X^{10'}\\
X^{11'}\\
\phi_0\\
\phi_1\\
\phi_2\\
\end{array}\right)=0
,
\ee
where $\chi_{A'a}, \psi_a,\phi_a$ are respectively the spinor components of $\chi_{E'ABCDE},\psi_{ABCD}$ and $\phi_{AB}$, following NP conventions. 
In a block-matrix form (\ref{m1}) can be written:
\be\label{m2}
\left(\begin{array}{cc}
M&N\\
\end{array}\right)\left(\begin{array}{c}
X\\
\phi\\
\end{array}\right)=0.\ee
These are complex equations. Including the complex conjugates in the system gives the square system
\be\label{m3}
\left(\begin{array}{ccc}
M&N&0\\
\overline{M}&0&\overline{N}
\end{array}\right)\left(\begin{array}{c}
X\\
\phi\\
\overline{\phi}\\
\end{array}\right)=0,\ee
so introduce the $10\times 10$ matrix $A$ (this matrix $A$ is conjugate to the previous $A$) by
\[A= \left(\begin{array}{ccc}
M&N&0\\
\overline{M}&0&\overline{N}
\end{array}\right).\]
Then the necessary condition for existence of a nontrivial solution to (\ref{c4}) is the vanishing of $\mbox{det}\,A$ and we turn next to this condition.

\medskip

We first deal with the case when the Weyl spinor is type N. In this case we may choose the spinor dyad so that $\psi_0\neq 0$ but $\psi_i=0$ otherwise. 
Then the matrix $N$ has one column of zeroes so that the rank of $A$ has dropped by at least two. In particular the determinant of $A$ is zero and (\ref{m3}) 
has nontrivial solutions. This is a particular instance of the problem with algebraically-special metrics noted above.

If the Weyl spinor is not of type N then we may choose the spinor dyad $(o^A,\iota^A)$ so that $\psi_0=0=\psi_4$, equivalently $o^A$ and $\iota^A$ are independent principal spinors of the Weyl spinor. 
Now we can row-reduce the block matrix $(M \;N)$ to
\[\left(\begin{array}{cccc}
*&4\psi_1&0&0\\
*&0&-2\psi_1&0\\
*&0&0&-2\psi_1\\
\tilde{r}_4&0&0&0\\
\tilde{r}_5&0&0&0\\
\end{array}\right)\]
where $\tilde{r}_4,\tilde{r}_5$ are combinations of the rows $r_1,\ldots r_5$ of $M$ which we'll give below. Now
\[\mbox{det}\,A=256\mbox{det}\left(\begin{array}{c}
\psi_1\tilde{r}_4\\
\psi_1^2\tilde{r}_5\\
\overline{\psi}_1\overline{\tilde{r}}_4\\
(\overline{\psi}_1)^2\overline{\tilde{r}}_5\\
\end{array}\right)\]
with
\be\label{r1}\psi_1\tilde{r}_4=\psi_1r_4-\frac{3}{2}\psi_2r_3+\psi_3r_2\ee
and
\be\label{r2}\psi_1^2\tilde{r}_5=\psi_1^2r_5-2\psi_1\psi_3r_3+\psi_3^2r_1.\ee
Our assumptions on the spinor dyad imply that the Weyl spinor can be written
\[\psi_{ABCD}=-4\psi_3o_{(A}o_Bo_C\iota_{D)}+6\psi_2o_{(A}o_B\iota_C\iota_{D)}-4\psi_1o_{(A}\iota_B\iota_C\iota_{D)}\]
from which we may calculate the gradients of the two scalar invariants:
\[\nabla_{AA'}I=2\psi^{BCDE}\nabla_{AA'}\psi_{BCDE}=2\psi^{BCDE}\chi_{A'ABCDE}\]
and
\[\nabla_{AA'}J=3\psi^{BCPQ}\psi^{DE}_{\;\;\;\;\;PQ}\chi_{A'ABCDE}.\]
Each can each be written as a pair of spinors by taking components on the index $A$. For $I$ we obtain 
\be\label{i1}\nabla_{0A'}I=4(-2\psi_3\chi_{A'1}+3\psi_2\chi_{A'2}-2\psi_1\chi_{A'3}),
\;\nabla_{1A'}I=4(-2\psi_3\chi_{A'2}+3\psi_2\chi_{A'3}-2\psi_1\chi_{A'4})\ee
while for $J$ it is simpler to write the combination
\be\label{j2}\nabla_{0A'}J+\frac{3}{2}\psi_2\nabla_{0A'}I=6(-\psi_3^2\chi_{A'0}+2\psi_1\psi_3\chi_{A'2}-\psi_1^2\chi_{A'4})\ee
and
\be\label{j3}\nabla_{1A'}J+\frac{3}{2}\psi_2\nabla_{1A'}I=6(-\psi_3^2\chi_{A'1}+2\psi_1\psi_3\chi_{A'3}-\psi_1^2\chi_{A'5}).\ee
Now comparing (\ref{r1})--(\ref{r2}) with (\ref{i1})--(\ref{j3}) and using (\ref{m1}) we see that $\mbox{det}\,A$ is a (nonzero) constant multiple of
\[\mbox{det}\left(\begin{array}{c}
dI\\
dJ\\
d\overline{I}\\
d\overline{J}\\
\end{array}\right),\]
where each row is written as a 4-component vector, so that the non-vanishing of $\mbox{det}\,A$ is indeed equivalent to (\ref{j1}). This completes the proof 
of Proposition 1.1.

As remarked above, in any algebraically-special space-time the Weyl spinor satisfies $I^3=6J^2$ so that (\ref{j1}) will not hold, and (\ref{c4}) 
admits solutions. More differentiation is needed to see whether these candidate KIDs are actual KIDs, 
but algebraically-special vacuum solutions without symmetries are known, for example some of the Robinson-Trautman solutions (see e.g. \cite{ExSo}). It 
remains a possibility that no algebraically-special metric without a symmetry can be asymptotically-flat\footnote{In this connection see \cite{ChrTod}.}.

Another way to see this last part of the Proposition, in the particular case that $J\neq 0$, is to contract (\ref{c4}) with $\chi^{BCDF}$. We obtain an expression for $\phi$, namely
\be\label{f1}\phi_A^{\;E}=\frac{1}{3}\chi^{BCDE}X^{E'F}\chi_{E'ABCDF}.\ee
Now contract (\ref{c4}) with $\psi^{ABCD}$ and $\psi^{ABPQ}\psi_{PQ}^{\;\;\;\;\;CD}$ respectively to obtain
\be\label{f2}X^a\nabla_aI=0=X^a\nabla_aJ.\ee
Since the system (\ref{f1})--(\ref{f2}) manifestly consists of five independent equations, it must exhaust (\ref{c4}), so that (\ref{c4}) is equivalent to 
defining $\phi_{AB}$ as in (\ref{f1}) and the two conditions (\ref{f2}) on $X^a$. Now it's clear that inconsistency of (\ref{c4}) must be (\ref{j1}).

\section{Rank of the Linear System}
We wish to confirm that the rank of the system in (\ref{m3}) is generically ten i.e. that generically $\mbox{det}\,A\neq 0$ or equivalently (\ref{j11}) holds. To this end, we follow \cite{bcs} and first claim that given any point $p$ there is a metric such that any value of the 
Riemann tensor and its derivative, consistent with symmetries, holds at $p$. Consider the metric
\[g_{ab}=\eta_{ab}-\frac{1}{3}t_{acbd}x^cx^d+\frac{1}{6}t_{eacbd}x^ex^cx^d,\]
in pseudo-Cartesian (or inertial) coordinates $x^a$, where $t_{acbd}$ and $t_{eacbd}$ are constant tensors with all the symmetries respectively of the Riemann tensor 
$R_{acbd}$ and its derivative $\nabla_eR_{acbd}$,
then at the origin of coordinates one readily calculates that
\[R_{abcd}=t_{abcd},\;\;\nabla_eR_{abcd}=t_{eabcd}.\]
Thus the Riemann tensor and its derivative at a point can take any value allowed by symmetry. 

We restrict to vacuum, so that the Ricci spinor and Ricci scalar are zero, and at a point $p$ we make the choices
\[\psi_0=\psi_2=\psi_4=0,\;\psi_1\psi_3\neq 0,\;\;|\psi_3|\neq 2|\psi_1|\]
for the Weyl spinor and
\[\chi_{A'0}=\chi_{A'3}=\chi_{A'4}=\chi_{A'5}=0,\;\;\chi_{A'1}=\iota_{A'},\;\chi_{A'2}=-io_{A'}\]
for its derivative. Then we may calculate
\[\nabla_aI=8\psi_3(i\ell_a+n_a),\;\;\nabla_aJ=6\psi_3(2i\psi_1\overline{m}_a-\psi_3m_a),\]
where $(\ell,n,m,\overline{m})$ form an NP-tetrad related in the conventional way to the normalised spinor dyad $(o^A,\iota^A)$ and its complex conjugate. Then
\[dI\wedge d\overline{I}\wedge dJ\wedge d\overline{J}=4608i|\psi_3|^4(|\psi_3|^2-4|\psi_1|^2)\ell\wedge n\wedge m\wedge\overline{m}\neq 0.\]
This shows that (\ref{j1}) is not constrained to fail by symmetry alone, so that we can expect it to hold generically.

\section*{Appendix}
The methods used here are also helpful for another problem considered in \cite{bcs}, namely the question of existence of a conformal Killing vector on a 
Riemannian 3-manifold $(\Sigma, g)$. We are led to 

\medskip

\noindent
{\bf{Proposition A.1}}:   {\it{
Given a Riemannian 3-manifold $(\Sigma, g)$, with Cotton-York tensor $Y_{ij}$ defined as in (\ref{cy1}), and the tensor $Z_{ij}$ as in (\ref{ev2}) then this metric admits no 
conformal Killing vectors if the three conformally invariant scalars $\tilde{\sigma}_2,\tilde{\sigma}_3,\tilde{\sigma}_4$ defined below as polynomial invariants of $Y_{ij}$ and $Z_{ij}$ are functionally independent:}}
\[d\tilde{\sigma}_2\wedge d\tilde{\sigma}_3\wedge d\tilde{\sigma}_4\neq 0.\]

Thus the sufficient condition for the nonexistence of conformal Killing vectors, (\ref{ev5}) below, is analogous to the condition (\ref{j1}) found above. 

The existence of conformal Killing vectors can only depend on the conformal class of $g$ and we should therefore expect conformal invariants and covariants to 
appear in the results. Recall the notion of conformal weight: under conformal rescaling
\[g_{ij}\rightarrow\hat{g}_{ij}=\Omega^2g_{ij}\]
we say that a geometrical quantity $\eta$, a tensor or a scalar, has conformal weight $w$ if
\[\eta\rightarrow\hat{\eta}=\Omega^w\eta.\]
A conformal Killing vector is a vector field $X^i$ such that
\[2D_{(i}X_{j)}:=\mathcal{L}_Xg_{ij} =\frac{2}{3}\phi g_{ij}\]
for some function $\phi$, which is evidently related to $X^i$ by $\phi=D_iX^i$. Note that $\phi$ transforms inhomogeneously under conformal rescaling, according to
\[\phi\rightarrow\hat{\phi}=\phi+3\mathcal{L}_X(\log\Omega).\]
We prolong as before, introducing a  bivector $F_{ij}$ so that
\be\label{ck1}D_iX_j=\frac{1}{3}\phi g_{ij}+F_{ij}.\ee
Then introduce the vector $V_i$ by
\be\label{ck2}D_i\phi=V_i,\ee
and commute derivatives on (\ref{ck1}) to obtain
\be\label{ck3}
D_iF_{jk}=-R_{ijk}^{\;\;\;\;\;\ell}X_\ell+\frac{1}{3}(V_jg_{ki}-V_kg_{ji}).\ee
Now commute derivatives on this to obtain
\be\label{ck4}D_iV_j=-3X^pD_pP_{ij}-2\phi P_{ij}-3P_{ip}F_j^{\;p}-3P_{kj}F_i^{\;p},\ee
where $P_{ij}$ is the Rho-tensor or Schouten tensor, defined in terms of the Ricci tensor and Ricci scalar by
\[P_{ij}:=R_{ij}-\frac{1}{4}Rg_{ij},\]
(and called $L_{ij}$ by \cite{bcs}). Recall that in dimension 3 the Riemann tensor is given in terms of the Rho-tensor by
\[R_{ijk\ell}=P_{ik}g_{j\ell}+P_{j\ell}g_{ik}-P_{jk}g_{i\ell}-P_{i\ell}g_{jk}.\]
Following \cite{bcs} we define the Cotton-York tensor by
\be\label{cy1}Y_{ij}=\epsilon_i^{\;km}D_mP_{kj},\ee
(called $H_{ij}$ by \cite{bcs}, also some authors have the opposite sign in this definition).
This tensor is trace-free, symmetric, divergence-free and has conformal weight $-1$, so that
\[Y_{ij}\rightarrow\hat{Y}_{ij}=\Omega^{-1}Y_{ij}\]
under conformal rescaling. It follows by an appeal to naturality that, for the conformal Killing vector $X^i$,
\be\label{cy2}\mathcal{L}_XY_{ij}=-\frac{1}{3}\phi Y_{ij},\ee
but we shall derive this equation below.

\medskip

The process of prolongation has led to a connection defined by (\ref{ck1})--(\ref{ck4}) on the vector bundle $\Lambda^0\oplus\Lambda^1\oplus\Lambda^1\oplus\Lambda^2$ with 
typical section $\psi_\alpha=(\phi,X_i,V_i,F_{ij})^T$. To construct the system we have already commuted derivatives on the first three of these so that the curvature of this connection is determined by commuting derivatives on the fourth. The result of doing this is then found to be precisely (\ref{cy2}), which can be written at length as
\be\label{cy3}X^pD_pY_{ij}+F_i^{\;p}Y_{jp}+F_j^{\;p}Y_{ip}+\phi Y_{ij}=0.\ee
This is the first set of necessary conditions on $\psi_\alpha$. The tensor on the LHS is symmetric and trace-free so that this represents five linear conditions on the ten components of $\psi_\alpha$. We need more, but first we make two deductions from (\ref{cy3}), by contracting with $Y^{ij}$ and $Y^{ik}Y^j_{\;\;k}$ respectively. The terms containing $F_{ij}$ drop out and we are left with
\be\label{ev1}X^pD_p(Y_{ij}Y^{ij})=-2\phi(Y_{ij}Y^{ij}),\;X^pD_p(Y_{ij}Y^{jk}Y_k^{\;i})=-3\phi(Y_{ij}Y^{jk}Y_k^{\;i}).\ee
Note that $\sigma_1:=Y_{ij}Y^{ij}$ and $\sigma_2:=Y_{ij}Y^{jk}Y_k^{\;i}$ are conformally weighted scalars with weights $-6$ and $-9$ respectively so that their combination $\tilde{\sigma}_2:=\sigma_2/(\sigma_1)^{3/2}$ is 
a conformally-invariant scalar which by (\ref{ev1}) is constant along $X^i$. (If $\sigma_1=0$ then $\Sigma$ is conformally flat and everything about conformal Killing vectors is known.)

Next we must differentiate (\ref{cy3}) to obtain more constraints. The divergence, or contracted derivative, automatically vanishes but the curl gives new information. First define the tensor
\be\label{ev2}
Z_{ij}=\epsilon_i^{\;mn}D_mY_{nj}.\
\ee
This is another trace-free, symmetric tensor but it is not conformally-weighted. In fact under conformal rescaling
\be\label{ev3}
Z_{ij}\rightarrow\hat{Z}_{ij}=\Omega^{-3}(Z_{ij}-2\epsilon_{(i}^{\;\;\;mn}Y_{j)n}\Upsilon_{m})\ee
with $\Upsilon_i=D_i(\log\Omega)$ as usual.

\medskip

Now the curl of (\ref{cy3}) is
\[\epsilon^{ikm} D_i(X^pD_pP_{kj}+F_k^{\;p}P_{jp}+F_j^{\;p}P_{kp}+\phi P_{kj}))\]
\[=X^pD_pZ_j^{\;m}+\frac{4}{3}\phi Z_j^{\;m}+F_j^{\;p}Z_p^{\;m}-F_p^{\;m}Z_j^{\;p}+\frac{2}{3}\epsilon^{ikm}V_iY_{kj}-\frac{1}{3}\epsilon_j^{\;km}Y_{kp}V^p.\]
Relabelling indices and algebraically manipulating the last two terms gives the second necessary condition:

\be\label{ev4}
X^pD_pZ_{ij}+\frac{4}{2}Z_{ij}+F_i^{\;k}Z_{jk}+F_j^{\;k}Z_{jk}+\frac{2}{3}\epsilon_{pq(i}V^pY_{j)}^{\;\;q}=0.\ee
The conditions (\ref{cy3}) and (\ref{ev4}) together give ten conditions  on the ten components of $\psi_\alpha$ so that we can expect an obstruction to the existence of a conformal Killing vector. To see  what it is, introduce the scalars
\[\sigma_3=Z^{ij}Y_{ij},\;\sigma_4=Z^{ij}Y_{ik}Y_j^{\;k}.\]
By (\ref{ev3}) these are conformally-weighted scalars with weights $-7$ and $-10$ respectively, and by (\ref{cy3}) and (\ref{ev4}) we may calculate
\[X^pD_p(\log\sigma_3)=-\frac{7}{3}\phi,\;\;X^pD_p(\log\sigma_4)=-\frac{10}{3}\phi.\]
We conclude that $\tilde{\sigma_3}:=\sigma_3/(\sigma_1)^{7/6}$ and $\tilde{\sigma}_4:=\sigma_4/(\sigma_1)^{5/3}$ are conformally-invariant scalars which are constant along $X^i$. Thus there can be no 
conformal Killing vector if the three conformally-invariant scalars that we have defined are functionally independent, that is if
\be\label{ev5}
d\tilde{\sigma}_2\wedge d\tilde{\sigma}_3\wedge d\tilde{\sigma}_4\neq 0,
\ee
which is our necessary condition.

The corresponding condition in \cite{bcs} is a polynomial in the tensors $Y_{ij},D_iY_{jk}$ and $D_iD_jY_{km}$, which is the same order of differentials as (\ref{ev5}).

\end{document}